\documentclass[aps,prd,preprint,superscriptaddress]{revtex4-2}
\usepackage{ulem}
\usepackage{xcolor}
\usepackage[colorlinks=true]{hyperref}
\usepackage{soul}
\usepackage{booktabs}
\usepackage{mathrsfs}
\usepackage{amssymb}

\usepackage{amsmath,amssymb,color,epsfig}
\allowdisplaybreaks[4]

\newcommand{\be}{\begin{equation}}
\newcommand{\ee}{\end{equation}}
\newcommand{\bea}{\setlength\arraycolsep{2pt} \begin{eqnarray}}
\newcommand{\eea}{\end{eqnarray}}
\newcommand{\nn}{\nonumber}

\def\0{{\sst{(0)}}}
\def\1{{\sst{(1)}}}
\def\2{{\sst{(2)}}}
\def\3{{\sst{(3)}}}
\def\4{{\sst{(4)}}}
\def\5{{\sst{(5)}}}
\def\6{{\sst{(6)}}}
\def\7{{\sst{(7)}}}
\def\8{{\sst{(8)}}}
\def\sst#1{{\scriptscriptstyle #1}}

\begin{document}

\hypersetup{
    linkcolor=blue,
    citecolor=red,
    urlcolor=magenta
}

% Use the \preprint command to place your local institutional report
% number in the upper righthand corner of the title page in preprint mode.
% Multiple \preprint commands are allowed.
% Use the 'preprintnumbers' class option to override journal defaults
% to display numbers if necessary
%\preprint{}

%Title of paper
\title{Solar-System Bounds on Ricci-flat Spindle Deformations of Schwarzschild}

% repeat the \author .. \affiliation  etc. as needed
% \email, \thanks, \homepage, \altaffiliation all apply to the current
% author. Explanatory text should go in the []'s, actual e-mail
% address or url should go in the {}'s for \email and \homepage.
% Please use the appropriate macro foreach each type of information

% \affiliation command applies to all authors since the last
% \affiliation command. The \affiliation command should follow the
% other information
% \affiliation can be followed by \email, \homepage, \thanks as well.

\author{Zhong-Xi Yu}
\email[]{zxy@jnnu.edu.cn}
\affiliation{College of Physics and Electronic Information Engineering, Jining Normal University, Wulanchabu, 012000, China}

\author{Hong-Da Lyu}
\email[]{hongdalyu@sdu.edu.cn}
\affiliation{Key Laboratory of Particle Physics and Particle Irradiation (MOE),
Institute of Frontier and Interdisciplinary Science,
Shandong University, Qingdao, Shandong, 266237, China}

\author{Shoulong Li}
\email[Corresponding author: ]{shoulongli@hunnu.edu.cn}
\affiliation{Department of Physics, Key Laboratory of Low Dimensional Quantum Structures and Quantum Control of Ministry of Education, and Institute of Interdisciplinary Studies, Hunan Normal University, Changsha, 410081, China}
\affiliation{Hunan Research Center of the Basic Discipline for Quantum Effects and Quantum Technologies, Hunan Normal University, Changsha 410081, China}

%Collaboration name if desired (requires use of superscriptaddress
%option in \documentclass). \noaffiliation is required (may also be
%used with the \author command).
%\collaboration can be followed by \email, \homepage, \thanks as well.
%\collaboration{}
%\noaffiliation

\date{\today}

\begin{abstract}

Recently, a new class of deformed black-hole exact solutions was constructed in four-dimensional general relativity. The deformation is controlled by a parameter $B$, which survives after demagnetizing a black hole immersed in an external Bertotti-Robinson magnetic field and changes the global structure of the spacetime into a non-asymptotically flat spindle geometry. Although no astrophysical mechanism for generating such a deformation is currently known, it is natural to ask phenomenologically how large such a geometric deformation could be if it extended over the weak-field solar exterior. Using two classical Solar-System tests, we derive the leading corrections to planetary perihelion precession and to the light travel time in a Shapiro-type configuration. Requiring the \(B\)-induced perihelion advance to be smaller than the observational uncertainties in the supplementary perihelion precessions of planets gives the strongest bounds, \( |B|\lesssim 10^{-24}\text{--}10^{-23}\ {\rm cm}^{-1}\), while a Cassini time delay estimate gives a complementary null-geodesic sensitivity at the level \( |B|\lesssim 10^{-21}\ {\rm cm}^{-1}\). These results show that any such spindle deformation, if extended to the solar exterior geometry, must be extremely suppressed on Solar-System scales.
  
\end{abstract}
% insert suggested keywords - APS authors don't need to do this
%\keywords{}

%\maketitle must follow title, authors, abstract, and keywords

\maketitle

\section{Introduction}

Exact solutions have always played a foundational role in general relativity, since they provide indispensable models for understanding nonlinear gravitational effects, spacetime structures, and the domain of validity of approximation schemes~\cite{Stephani:2003tm,Griffiths:2009dfa}. Recently, Astorino~\cite{Astorino:2026okd} constructed a novel static black-hole solution in four-dimensional general relativity by applying a solution-generating technique to a Schwarzschild black hole immersed in an external Bertotti-Robinson magnetic field~\cite{Ernst:1976mzr}. The strength of the seed magnetic field is characterized by a parameter \(B\), and the solution is generated through symmetry transformations of the Ernst equations~\cite{Astorino:2025lih}. Remarkably, after the magnetic field is removed, \(B\) survives as a remnant in the resulting Ricci-flat metric and was interpreted as a new type of geometric ``hair'', deforming the event horizon into an oblate spheroid configuration and rendering the spacetime no longer asymptotically flat.

More recently, Ma and L\"u~\cite{Ma:2026otg} obtained a rotating counterpart by demagnetizing~\cite{Gibbons:2013yq} the Kerr--Bertotti--Robinson solution~\cite{Podolsky:2025tle,Ovcharenko:2025cpm}. Their Ricci-flat rotating metric represents a deformation of the Kerr spacetime characterized by the same type of parameter $B$, where the standard asymptotic Kerr region is replaced by a regular spindle-shaped dome. The resulting geometry was referred to as a deformed ``$B$-spindle'', emphasizing that the parameter $B$ controls an intrinsic spindle deformation of the black-hole spacetime.

These pioneering developments open an interesting new avenue in the study of exact black-hole spacetimes in four-dimensional general relativity, revealing an unexpected class of Ricci-flat but non-asymptotically flat geometries. It has been shown that the parameter \(B\) modifies the expressions of the conserved charges, such as the mass and angular momentum, as well as the temperature, while the thermodynamic first law can still be consistently established~\cite{Astorino:2026okd,Ma:2026otg}. Meanwhile, the global structure and physical interpretation of related symmetry-transformed solutions have also been discussed recently~\cite{Herdeiro:2026jem}. These results indicate that \(B\) is an integration constant with genuine geometric and thermodynamic effects. While clarifying the physical origin of \(B\) is important, it is still meaningful to adopt a phenomenological viewpoint and, without specifying how an astrophysical object could acquire such a \(B\)-deformation, ask how strongly it would be constrained if it were present in the weak-field solar exterior.

In this work, we address this question for the static Ricci-flat spindle deformation of Schwarzschild. We focus on two precision Solar-System tests. The first is the anomalous perihelion precession of planets, which probes timelike geodesics. The second is the Shapiro time delay, which probes null geodesics. For both cases, we work perturbatively in \(B^2\) and derive the leading corrections in the weak-field regime. By comparing these corrections with observational residuals from planetary ephemerides and with a Cassini-inspired finite-distance light-time estimate, we obtain phenomenological Solar-System bounds on the deformation parameter \(B\). These bounds provide a quantitative assessment of the allowed size of the deformation in the weak-field regime and help clarify whether this novel class of Ricci-flat but non-asymptotically flat solutions can have observable relevance on Solar-System scales.

The remainder of this paper is organized as follows. In Sec.~\ref{spindle}, we briefly review the static Ricci-flat spindle deformation of Schwarzschild and the relation between the integration constant and the physical mass. In Sec.~\ref{solar}, we derive the Solar-System bounds on $B$ from planetary perihelion precessions and from a Shapiro-type light-time constraint. Finally, Sec.~\ref{conclusion} summarizes the results and presents the outlook.

\section{Spindle-deformed Schwarzschild spacetime} \label{spindle}

In this section, we briefly review the newly found static solution of general relativity obtained in Ref.~\cite{Astorino:2026okd}. The metric can be written as
\begin{equation}
ds^2 = \frac{(1+\Omega+B^2 m r x^2)^2}{4\Omega^4}\left[ -\frac{Q\sigma^2}{r^2 } dt^2 + \frac{r^2}{Q} dr^2 + \frac{r^2 dx^2}{(1-x^2) P} \right]  + \frac{4 r^2 (1-x^2 ) P d\phi^2}{P_0^2 (1+\Omega+B^2 m r x^2)^2}   \,,  \label{metric}  
\end{equation}
where \(x=\cos\theta\). The metric functions $P$, $Q$, $\Omega$ and $\Delta$ are given by
\bea
P &=& 1+B^2 m^2 x^2 \,, \nn \\ 
Q &=& (1+B^2 r^2)\Delta \,,  \nn \\ 
\Omega^2 &=& 1+B^2 r^2 - B^2 x^2 \Delta \,,  \nn \\ 
\Delta &=&  (1-B^2 m ^2) r^2 -2 m r  \,.
\eea
Here \(m\) is an integration constant which reduces to the Schwarzschild mass in the limit \(B\to0\), while \(B\) characterizes the spindle deformation of the Schwarzschild geometry. The parameter \(P_0\) is fixed by requiring the azimuthal coordinate \(\phi\) to have the standard period \(2\pi\), thereby ensuring the regularity of the symmetry axis. The parameter \(\sigma\) is introduced as a normalization of the timelike Killing vector such that the physical mass of the black hole obeys the standard thermodynamic first law. The parameters \(P_0\) and \(\sigma\) are given by
\be
P_0 = \sigma^{-2} = (1+B^2 m ^2) \,.
\ee
The solution~\eqref{metric} can also be viewed as the static limit of the rotating metric reported in Ref.~\cite{Ma:2026otg}.

The physical mass associated with the metric~\eqref{metric} is obtained in Ref.~\cite{Astorino:2026okd} as
\be
M =  \frac{m}{(1+B^2 m ^2)^{3/2}} \,. \label{mass}
\ee
Similarly, the Hawking temperature and the Bekenstein-Hawking entropy are given by~\cite{Astorino:2026okd}
\be
T =  \frac{(1+B^2 m ^2)^{3/2}}{8\pi m} \,,\quad S = \frac{4 \pi  m^2}{\left(1 +B^2 m^2\right)^3} \,.\nn
\ee
Keeping the deformation parameter \(B\) fixed, these thermodynamic quantities satisfy the differential first law and the Smarr relation,
\be
d M = T d S \,, \quad   M = 2 T S \,.\nn
\ee
Therefore, \(M\), rather than the integration constant \(m\), should be identified as the physical mass entering the thermodynamic description of the black hole. It is useful to invert Eq.~\eqref{mass} and express the integration constant \(m\) in terms of the physical mass \(M\). Solving Eq.~\eqref{mass} gives
\be
m^2 = \frac{2 \sqrt{3}}{3 B^3 M} \cos \left\{\frac{2 \pi}{3}-\frac{1}{3} \cos ^{-1}\bigg[-\frac{3 \sqrt{3}}{2}  B M \bigg]\right\} -\frac{1}{B^2} \,,\label{massrelation}
\ee
The branch in Eq.~\eqref{massrelation} is chosen such that the Schwarzschild limit is recovered smoothly, namely \(m\to M\) as \(B\to0\). For a small deformation parameter \(B\), Eq.~\eqref{massrelation} can be expanded as
\be
m = M +\frac{3 B^2 M^3}{2 } + {\cal O}(B^4) \,.
\ee
This expansion shows that the difference between the integration constant \(m\) and the physical mass \(M\) starts at order \(B^2\). Therefore, when deriving the leading \(B^2\)-corrections to Solar-System observables, this difference should be included consistently. Substituting Eq.~\eqref{massrelation} into the metric~\eqref{metric}, the spacetime can be parameterized by the physical mass \(M\) and the deformation parameter \(B\). In the Solar-System application below, \(M\) is identified with the usual solar gravitational mass in the Schwarzschild limit, while the \(B\)-dependent terms are treated as additional geometric corrections to be constrained. We also note that \(B\) is not simply a correction tied only to the central mass: when \(m=0\), the spacetime~\eqref{metric} does not reduce to Minkowski spacetime, but becomes a warped product of \(U(1)\) and \({\rm AdS}_3\)~\cite{Astorino:2026okd,Ma:2026otg}. This reflects the fact that \(B\) also characterizes the large-distance background structure of the solution.

\section{Solar-System Constraints} \label{solar}

In this section, we derive observational bounds on the spindle deformation parameter \(B\) from two classical high-precision tests in the Solar System: the anomalous perihelion precession of planets and the Shapiro time delay. The former probes the motion of massive bodies, while the latter is associated with the propagation of light rays. Accordingly, we first summarize the geodesic motion of test particles in the background spacetime~\eqref{metric}.

Before doing so, we clarify the approximation used in the following geodesic analysis. Since the metric~\eqref{metric} is invariant under the reflection \(x\to -x\), the equatorial plane \(x=0\) is preserved by the geodesic motion. In what follows, we restrict to this reflection-symmetric equatorial sector by imposing \(x=0\) and \(\dot{x}=0\). The bounds obtained below should therefore be understood as phenomenological bounds within this equatorial-plane approximation. The trajectory of a test particle is denoted by
\begin{equation}
 X^\mu(\lambda)=\bigl(t(\lambda),r(\lambda),x(\lambda),\phi(\lambda)\bigr),
\end{equation}
where \(\lambda\) is an affine parameter along the geodesic. The tangent vector \(U^\mu={dX^\mu}/{ d\lambda}\) satisfies the geodesic equation
\begin{equation}
 \ddot X^\mu+\Gamma^\mu{}_{\rho\sigma}\dot X^\rho\dot X^\sigma=0 \,,
\end{equation}
together with the normalization condition
\begin{equation}
 g_{\rho\sigma}\dot X^\rho\dot X^\sigma=-\epsilon \,.
 \label{norm_condition}
\end{equation}
Here \(\Gamma^\mu{}_{\rho\sigma}\) denotes the affine connection associated with the metric \(g_{\mu\nu}\), and a dot denotes differentiation with respect to \(\lambda\). The parameter \(\epsilon\) distinguishes timelike and null geodesics: \(\epsilon=1\) corresponds to timelike geodesics, while \(\epsilon=0\) corresponds to null geodesics. For timelike geodesics, the affine parameter \(\lambda\) can be chosen as the proper time \(\tau\). Since the metric~\eqref{metric} is independent of \(t\) and \(\phi\), the spacetime admits two Killing vectors,
\begin{equation}
 \xi^\mu_{(t)}=(\partial_t)^\mu,\qquad
 \xi^\mu_{(\phi)}=(\partial_\phi)^\mu \,. \nn
\end{equation}
For any affinely parametrized geodesic, the contractions of the tangent vector with these Killing vectors are conserved along the trajectory. We therefore define
\begin{equation}
  E\equiv -g_{\mu\nu}\xi^\mu_{(t)}\dot X^\nu,
 \qquad
  L\equiv g_{\mu\nu}\xi^\mu_{(\phi)}\dot X^\nu .
 \label{general_EL}
\end{equation}
For the static metric~\eqref{metric}, these conserved quantities reduce to
\begin{equation}
  E=-g_{tt}\dot t,\qquad
  L=g_{\phi\phi}\dot\phi .
 \label{EL_general}
\end{equation}
For timelike geodesics, \(E\) and \(L\) can be interpreted as the energy and angular momentum per unit rest mass. For null geodesics, they are conserved quantities associated with time-translation and axial symmetries, and their ratio \(L/E\) plays the role of an impact parameter in the weak-field limit. These relations will be used below to derive the corrections to planetary perihelion precession and to the Shapiro time delay.

\subsection{Perihelion precession} \label{solar1}

We first apply the general geodesic relations derived above to the motion of massive planets. For timelike geodesics, we set \(\epsilon=1\) in Eq.~\eqref{norm_condition} and choose the affine parameter as the proper time \(\tau\). The normalization condition then becomes
\be
 g_{tt} \dot{t}^2 +g_{rr} \dot{r}^2 +g_{xx} \dot{x}^2 +g_{\phi\phi} \dot{\phi}^2 = -1 \,, \label{orbiteq}
\ee
where a dot denotes differentiation with respect to \(\tau\). Specializing to the equatorial plane \(x=0\), and using \(\dot{x}=0\), one can solve Eq.~\eqref{EL_general} for \(\dot t\) and \(\dot\phi\). Substituting the results into Eq.~\eqref{orbiteq}, one obtains
\be
\dot{r}^2 +\frac{1}{g_{rr}} +\frac{{ E}^2}{g_{tt} g_{rr}}  +\frac{{ L}^2}{g_{rr} g_{\phi\phi}}= 0 \,.\label{radial_timelike}
\ee
It is convenient to introduce the inverse radial variable \(u=r^{-1}\) and use the azimuthal angle \(\phi\) to parametrize the orbit. After rewriting the radial equation~\eqref{radial_timelike} in terms of \(u\), differentiating the resulting first-order orbital equation with respect to \(\phi\) yields the second-order orbit equation
\be
u''+u-\frac{M}{{L}^2} = 3Mu^2 + F(u) B^2 +{\cal O}(B^4)  \,, \label{orbit_eq_B}
\ee
where we have used the relation~\eqref{massrelation} between the integration constant \(m\) and the physical mass \(M\), and then expanded the result for small \(B\). A prime denotes differentiation with respect to \(\phi\). The function \(F(u)\) collects the leading \(B\)-dependent contribution induced by the spindle deformation and is given by
\be
F= 
\frac{3-4E^2}{2L^2u^3}
-\frac{3M}{2L^2u^2}
+\frac{4ML^2-5M^3}{2L^2}
+3M^2u
-\frac{3M^3u^2}{2} \,.
\ee
To solve Eq.~\eqref{orbit_eq_B}, we decompose the orbit as \(u = u_0 + u_{\rm Sch} + u_{B}\), where \(u_0\) is the Newtonian solution, \(u_{\rm Sch}\) denotes the standard Schwarzschild correction, and \(u_B\) represents the correction induced by the spindle deformation. In the limit \(B\to0\), Eq.~\eqref{orbit_eq_B} reduces to the standard Schwarzschild result. At Newtonian order, one has
\be
u_0''+u_0-\frac{M}{L^2}=0 .
\label{u0_equation}
\ee
The corresponding Keplerian orbit is
\be
u_0 = \frac{1}{p} (1 + e \cos \phi) \,,\quad {\rm with} \quad {p} = \frac{L^2}{M}  \,,
\ee
where \(e\) is the orbital eccentricity and \(p\) is the semi-latus rectum.
At the next order, the Schwarzschild correction and the \(B^2\)-correction satisfy
\bea
u_{\rm Sch}''+u_{\rm Sch} &=& 3Mu_0^2 \,, \label{uSch_eq} \\
u_{B}''+u_{B} &=& F(u_0) B^2 \,. \label{uB_eq}
\eea
The particular solution of Eq.~\eqref{uSch_eq} containing the secular term is
\be
u_{\rm Sch} = \frac{3M}{p^2}\left(1+\frac{e^2}{2} - \frac{e^2}{6} \cos 2\phi + e \phi  \sin\phi \right)\,. \label{uSch_solution}
\ee
Therefore, up to nonsecular terms, the Schwarzschild-corrected orbit can be written as
\be
u_0+u_{\rm Sch}
\simeq
\frac{1}{p}
\Bigg\{
1
+\frac{3M}{p}
\left[
1+\frac{e^2}{2}
-\frac{e^2}{6}\cos 2\phi
\right]
+
e\cos\left[
\left(1-\frac{3M}{p}\right)\phi
\right]
\Bigg\}.
\label{Sch_orbit_shift}
\ee
This gives the standard Schwarzschild perihelion advance
\be
\Delta\phi_{\rm Sch}
=
\frac{6\pi M}{p} \label{deltaphisch}
\ee
per orbital period. The secular term proportional to \(\phi\sin\phi\) in \(u_{\rm Sch}\) is responsible for the perihelion advance, while the remaining bounded terms only modify the shape of the orbit without producing a cumulative shift. The same reasoning applies to the correction induced by the \(B\) deformation. Therefore, in order to extract the perihelion shift generated by \(B\), it is sufficient to identify the resonant part of the source term \(F(u_0)\). We write
\begin{equation}
 F(u_0)=C\cos\phi+\hbox{non-resonant terms},
\end{equation}
where the resonant coefficient is extracted by orthogonality:
\begin{equation}
 C={1\over \pi}\int_0^{2\pi} F(u_0)\cos\phi\,d\phi \,.
\end{equation}
Thus the resonant equation~\eqref{uB_eq} has the secular particular solution
\begin{equation}
 u_B={C B^2\over 2}\phi\sin\phi + \dots\,,
\end{equation}
where
\be
C = %\left[
 {3e(4E^2-3)p^2\over 2M (1-e^2)^{5/2}}
 +{3ep\over (1-e^2)^{3/2}}
 +{3eM^2\over p}
 -{3eM^3\over p^2} \,. \label{C_result}
% \right].
\ee
This secular term can be absorbed into a small shift of the orbital frequency because
\begin{equation}
u = \frac1p \left\{\alpha +e\cos\left[\left(1 -\frac{3M}{p} -\frac{Cp B^2}{2e}\right)\phi\right]\right\} \,.
\end{equation}
where \(\alpha\) contains the nonsecular modifications of the orbital shape induced by \(B\). The corresponding perihelion advance induced by the spindle deformation is
\be
\Delta \phi_B = \frac{\pi C p B^2}{ e} \,.
\ee
For Solar-System planets, \(p/M\gg1\), so the leading contribution to the resonant coefficient in Eq.~\eqref{C_result} comes from the term proportional to \(p^2\). In addition, \(E\simeq1\) for weak-field planetary motion, with deviations suppressed by \(M/p\). The leading \(B\)-induced perihelion shift is therefore
\be
\Delta \phi_B
\simeq
\frac{3\pi B^2 c^2 p^3}
{2GM(1-e^2)^{5/2}} 
\label{Delta_phi_B}
\ee
per orbital period, where we have restored the factors of \(G\) and \(c\). We use centimeter-gram-second (CGS) units in the numerical analysis.

As an illustration, let us consider Mercury. Its orbital parameters are~\cite{planetdata}
\bea
 a&=&5.7909\times 10^{12}\,{\rm cm} \,,
\nn\\
 e&=&0.20563 \,,
\nn\\
 p&=&a(1-e^2)=5.5460\times 10^{12}\,{\rm cm} \,.\nn
\label{Mercury_parameters}
\eea
For Mercury one has \(pc^2/(G M_\odot)\simeq 3.8\times10^7\). 
Here \(a\), \(e\), and \(p\) denote the semi-major axis, eccentricity, and semi-latus rectum of the orbit, respectively. The standard Schwarzschild contribution, calculated from Eq.~\eqref{deltaphisch}, is \(N\Delta\phi_{\rm Sch} =42.98''/{\rm century}\), where \(N\simeq415.2\) is the number of Mercury orbits per century. The supplementary perihelion precession of Mercury is \(-0.0020\pm0.0030\ ''/{\rm century}\)~\cite{Pitjeva:2013xxa}. Since the leading \(B\)-induced correction is proportional to \(B^2\), we conservatively require its magnitude to be smaller than the quoted observational uncertainty. Thus, for Mercury we impose
$
|N\Delta\phi_B|
\lesssim
0.0030''/{\rm century},
$
where $N\simeq415.2$ is the number of Mercury orbits per century. This gives
\be
|B|
\lesssim
7.6\times10^{-23}\ {\rm cm}^{-1},
\qquad
|B|^{-1}
\gtrsim
1.3\times10^{22}\ {\rm cm}.
\ee
Repeating the same analysis for other Solar-System planets gives the phenomenological bounds listed in Table~\ref{tab:perihelion_bounds}. The perihelion-precession bounds on \(B\) are at the level of \(10^{-24}\text{--}10^{-23}\,{\rm cm}^{-1}\).
\begin{table*}[t]
\centering
\begin{tabular}{cccccc}
\hline\hline
$\quad$Planet$\quad$
& $\quad$\(a\) \(({\rm AU})\)~\cite{planetdata}$\quad$
& $\quad$\(e\)~\cite{planetdata}$\quad$
& $\quad$\(p\) \((10^{12}{\rm cm})\)$\quad$
& $\quad$\(\dot\varpi_{\rm supp}\) \(({}''/{\rm century})\)~\cite{Pitjeva:2013xxa}$\quad$
& $\quad$\(|B|_{\rm max}\) \(({\rm cm}^{-1})\)$\quad$
\\
\hline
Mercury
& 0.38709927
& 0.20564
& 5.5460
& \(-0.0020\pm0.0030\)
& \(7.6\times10^{-23}\)
\\
Venus
& 0.72333566
& 0.00678
& 10.8205
& \(0.0026\pm0.0016\)
& \(3.4\times10^{-23}\)
\\
Earth
& 1.00000261
& 0.01671
& 14.9556
& \(0.00019\pm0.00019\)
& \(9.3\times10^{-24}\)
\\
Mars
& 1.52371034
& 0.09339
& 22.5956
& \(-0.000020\pm0.000037\)
& \(3.0\times10^{-24}\)
\\
Jupiter
& 5.20288700
& 0.04839
& 77.6519
& \(0.0587\pm0.0283\)
& \(3.3\times10^{-23}\)
\\
Saturn
& 9.53667594
& 0.05386
& 142.2528
& \(-0.00032\pm0.00047\)
& \(2.7\times10^{-24}\)
\\
\hline\hline
\end{tabular}
\caption{Phenomenological bounds on the spindle deformation parameter \(B\) obtained from the supplementary perihelion precessions of Solar-System planets. The bounds are derived by requiring the magnitude of the \(B\)-induced perihelion advance to be smaller than the quoted observational uncertainty in \(\dot\varpi_{\rm supp}\). For Earth, the Earth-Moon barycenter data are used.}
\label{tab:perihelion_bounds}
\end{table*}

\subsection{Shapiro time delay} 

We next consider a Shapiro-type light-time configuration~\cite{Shapiro:1964uw}. A radar signal is emitted from the Earth, propagates near the Sun, reaches a target planet or spacecraft, and is then reflected back to the Earth. In the Schwarzschild geometry, the propagation time is larger than that in flat spacetime, giving the standard Shapiro time delay. In the present spacetime, the light travel time also receives corrections from the spindle deformation parameter \(B\). 

The propagation of the radar signal is described by a null geodesic. Therefore, we set \(\epsilon=0\) in Eq.~\eqref{norm_condition}. The normalization condition becomes
\be
 g_{tt} \dot{t}^2 +g_{rr} \dot{r}^2 +g_{xx} \dot{x}^2 +g_{\phi\phi} \dot{\phi}^2 = 0 \,, \label{orbiteqnull}
\ee
where a dot denotes differentiation with respect to the affine parameter \(\lambda\). As in the timelike case, we work in the equatorial sector \(x=0\) with \(\dot{x}=0\). The conserved quantities associated with the Killing vectors \(\partial_t\) and \(\partial_\phi\) are still given by Eq.~\eqref{EL_general}, although for null geodesics they should be understood as conserved quantities along the ray. Solving these relations for \(\dot t\) and \(\dot\phi\), and substituting them into Eq.~\eqref{orbiteqnull}, one obtains
\be
\dot{r}^2 +\frac{{ E}^2}{g_{tt} g_{rr}}  +\frac{{ L}^2}{g_{rr} g_{\phi\phi}}= 0 \,.\label{radial_null}
\ee
To compute the propagation time of the radar signal, it is useful to use the coordinate time \(t\), rather than the affine parameter \(\lambda\), to parametrize the null trajectory. From \(\dot{r}=\dot{t}\,dr/dt\), the radial null equation~\eqref{radial_null} can be rewritten as 
\be
\left(\frac{dr}{dt}\right)^2 = -\frac{g_{tt}}{g_{rr}} - \frac{g_{tt}^2 L^2}{g_{rr}g_{\phi\phi}E^2} \label{drdteq}
\ee
At the point of closest approach to the Sun, denoted by \(r=b\), one has \(dr/dt=0\). This gives
\be
\frac{L^2}{E^2} = -\frac{g_{\phi\phi}(b)}{g_{tt}(b)} \,. \label{impact_relation}
\ee 
Here \(b\) denotes the distance of closest approach of the light ray to the Sun, which plays the role of the impact parameter in the weak-field Solar-System setup. Substituting Eq.~\eqref{impact_relation} into Eq.~\eqref{drdteq}, and solving for \(dt/dr\), we obtain
\bea
\frac{dt}{dr} &=& \sqrt{-\frac{g_{rr}}{g_{tt}}} \left( 1 - \frac{g_{\phi\phi}(b)}{g_{tt}(b)}\frac{g_{tt}}{g_{\phi\phi}} \right)^{-\frac12} \nn \\
&=& \left[\frac{1}{1-\frac{2 M}{r}} +\frac{B^2 r^2 \left(\frac{4 M^3}{r^3}+\frac{3 M^2}{r^2}+\frac{4 M}{r}-2\right)}{2 \left(1-\frac{2 M}{r}\right)^2} \right]  \left[ 1 - \frac{b^2 \left(1-\frac{2 M}{r}\right)}{r^2 \left(1-\frac{2 M}{b}\right)} - \frac{5 b B^2 M^3 (r-b)}{r^3 \left(1-\frac{2 M}{b}\right)^2}\right]^{-\frac12} \nn \\
&=& \frac{\left(1-\frac{2 M}{r}\right)^{-1} }{\left[1-\frac{b^2 \left(1-\frac{2 M}{r}\right)}{r^2 \left(1-\frac{2 M}{b}\right)}\right]^{\frac12}} +\frac{\frac{r \left(4 M^3+3 M^2 r+4 M r^2-2 r^3\right) \left(1-\frac{b^2 \left(1-\frac{2 M}{r}\right)}{r^2 \left(1-\frac{2 M}{b}\right)}\right)}{(2 M-r)^2}-\frac{5 b^3 \left(b M^3-M^3 r\right)}{r^3 (2 M-b)^2 \left(1-\frac{2 M}{r}\right)}}{2 \left(1-\frac{b^2 \left(1-\frac{2 M}{r}\right)}{r^2 \left(1-\frac{2 M}{b}\right)}\right)^{3/2}} B^2 \nn \\
&=& \sqrt{\frac{r^2}{r^2-b^2}} \left(1 +\frac{2 M}{r} +\frac{b M}{r (b+r)}\right) +\frac{2 B^2 M r^2}{\sqrt{r^2-b^2}}+\frac{b B^2 M r^2 (r-b)}{\left(r^2-b^2\right)^{3/2}} -\frac{B^2 r^3}{\sqrt{r^2-b^2}}
\eea
In the second and third equalities, the integration constant \(m\) has been replaced by the physical mass \(M\), and the result has been expanded to the leading nontrivial order in the deformation parameter, \({\cal O}(B^2)\). In the last equality, we have further expanded in the weak-field regime \(M/r\ll1\) and \(M/b\ll1\), keeping the leading contributions relevant for the Solar-System light-time effect.

The correction to the propagation time, relative to the flat-spacetime propagation from the point of closest approach \(b\) to a radial position \(r_0\), is approximately given by
\be
\Delta T(r_0) = \int dt = ( T_{\rm Sch}(r) -T_0(r)+  T_{B}(r))|_{r=b}^{r=r_0}  \,,
\ee
where \(T_0(r)\) denotes the flat-spacetime propagation time, \(T_{\rm Sch}(r)\) denotes the propagation time in the Schwarzschild geometry, and \(T_B(r)\) is the correction induced by the \(B\) deformation. Explicitly, one has
\bea
T_{B} &=& \frac{B^2 \sqrt{r^2-b^2}}{3}   \left[\frac{3 b^2 M}{b+r}-2 b^2+3 M (b + r)-r^2\right] \,, \nn \\
 T_{\rm Sch} &=& T_0 +2 M \ln (r + \sqrt{r^2 -b^2}) + M \sqrt{\frac{r - b}{r +b}} \,, \nn \\
T_0 &=& \sqrt{r^2 - b^2} \,.
\eea
For a radar signal emitted from the Earth, reflected by a target body, and received back on Earth, the total round-trip correction is
\be
\Delta T_{\rm round}
=
2\left[
\Delta T(r_E)+\Delta T(r_R)
\right],
\label{round_trip_delay}
\ee
where \(r_E\) and \(r_R\) denote the heliocentric radial distances of the Earth and the reflecting body, respectively. When the radar signal grazes the solar surface, the standard Shapiro effect is maximized. In this case the distance of closest approach can be approximated by the solar radius, \(b\simeq R_\odot\). Since \(r_E,r_R\gg b\), the standard Schwarzschild time delay becomes
\be
\Delta T_{\rm Sch}^{\rm round} = \frac{4 G M}{c^3} \left[ \ln \left( \frac{4 r_E r_R}{b^2} \right) +1 \right] \,,
\ee
while the leading \(B\)-dependent contribution to the round-trip light-time correction is
\be
\Delta T_{B}^{\rm round}
\simeq
-\frac{2B^2}{3c}
\left(
r_E^3+r_R^3
\right),
\label{round_trip_B_leading}
\ee
where we have also restored the factors of \(G\) and \(c\). The leading term in Eq.~\eqref{round_trip_B_leading} is independent of \(M\). This is not unexpected, because \(B\) also characterizes the non-asymptotically flat background structure of the solution. Therefore, this term should not be interpreted as an ordinary Shapiro delay sourced only by the solar mass. Rather, it is a finite-distance light-time correction induced by the \(B\)-deformed background geometry. 

The Shapiro time delay is usually used to constrain the parameterized post-Newtonian (PPN) parameter \(\gamma\). In the standard PPN setting, the round-trip time delay can be written as
\be
\Delta T_{\rm PPN}^{\rm round}
=
\frac{(1+\gamma)}{2}
\Delta T_{\rm Sch}^{\rm round} \,,
\ee
where \(\gamma=1\) gives the Schwarzschild result. Since the present spacetime is not asymptotically flat and the \(B\)-dependent correction has a different distance dependence from the PPN Shapiro delay, the Cassini constraint on \(\gamma\) cannot be regarded as a direct PPN bound on \(B\). We therefore use it only as a phenomenological reference for estimating the allowed size of the finite-distance light-time correction and impose
\be
\left|
\frac{\Delta T_{B}^{\rm round}}
{\Delta T_{\rm Sch}^{\rm round}}
\right|
\lesssim
\frac{|\gamma-1|}{2}.
\label{gamma_proxy}
\ee
For a Cassini-inspired solar-conjunction estimate, we take the heliocentric distance of the spacecraft to be \(r_R=8.43\,{\rm AU}\), the distance of closest approach to be \(b=1.6R_\odot\), and \(r_E=1\,{\rm AU}\). Using the Cassini result \(\gamma-1=(2.1\pm2.3)\times10^{-5}\)~\cite{Bertotti:2003rm} as the reference sensitivity, we take \(|\gamma-1|\lesssim2.3\times10^{-5}\). This gives the estimate
\be
|B|
\lesssim
8.5\times10^{-21}\ {\rm cm}^{-1},
\qquad
|B|^{-1}
\gtrsim
1.2\times10^{20}\ {\rm cm}.
\ee
This estimate is weaker than the perihelion-precession constraints obtained above, but it provides an independent null-geodesic sensitivity to the spindle deformation. For the value of \(B\) allowed by Cassini, one has \(|B|r_R\sim 10^{-6}\), so the perturbative condition \(|B|r\ll1\) is self-consistently satisfied along the light path.

\section{Conclusion} \label{conclusion}

Recently, a new class of four-dimensional Ricci-flat black-hole solutions in general relativity has been proposed, characterized by a spindle deformation parameter \(B\). This parameter deforms the spacetime geometry and renders the solution non-asymptotically flat, while preserving Ricci flatness. Since such a deformation arises in an exact solution of general relativity, it is meaningful to ask how strongly it can be constrained by classical observations. In this work, we have placed phenomenological Solar-System bounds on the spindle deformation parameter \(B\) by using two precision tests: the anomalous perihelion precession of planets and the Shapiro time delay. We found that planetary perihelion precessions constrain the spindle deformation parameter to the level \( |B|\lesssim 10^{-24}\text{--}10^{-23}\ {\rm cm}^{-1}\), while a Cassini-inspired finite-distance light-time estimate gives a complementary null-geodesic sensitivity at the level \( |B|\lesssim 10^{-21}\ {\rm cm}^{-1}\). These results show that, if such a \(B\)-deformation were present over the weak-field solar exterior, its effective value would have to be extremely small on Solar-System scales.

Nevertheless, independently of whether this metric can arise as the exterior geometry of realistic astrophysical objects, the exact black-hole solution itself remains of theoretical interest. In particular, studying how the parameter \(B\) modifies the near-horizon spacetime structure can help clarify the geometric role of the spindle deformation in general relativity. It would therefore be interesting to investigate its effects on strong-field properties such as circular geodesics, photon spheres, black-hole shadows, and gravitational lensing. These questions are left for future work~\cite{Lyu2026}.

\begin{acknowledgments}

We are grateful to H. L\"u, and Liang Ma for useful discussions. 
S.L. is supported in part by the National Natural Science Foundation of China (No. 12105098, No. 12481540179) and the Natural Science Foundation of Hunan Province (No. 2022JJ40264), and the innovative research group of Hunan Province under Grant No. 2024JJ1006, and by the Excellent Young Scholars Program of the Hunan Provincial Department of Education under Grant No. 25B0092. 
H.D.L.~is  supported in part by Postdoctoral Innovation Project of Shandong Province SDCX-ZG-202503036 and National Natural Science Foundation of China Grants No.12447134. 
Z.X.Y is supported in part by the Young Scholars Startup Fund of Jining Normal University.

\end{acknowledgments}

% Create the reference section using BibTeX:
%\bibliography{basename of .bib file}

\begin{thebibliography}{99}

%\cite{Stephani:2003tm}
\bibitem{Stephani:2003tm}
H.~Stephani, D.~Kramer, M.~A.~H.~MacCallum, C.~Hoenselaers and E.~Herlt,
``Exact solutions of Einstein's field equations,''
Cambridge Univ. Press, 2003,
ISBN 978-0-521-46702-5, 978-0-511-05917-9
%doi:10.1017/CBO9780511535185
%798 citations counted in INSPIRE as of 06 Jun 2026

%\cite{Griffiths:2009dfa}
\bibitem{Griffiths:2009dfa}
J.~B.~Griffiths and J.~Podolsky,
``Exact Space-Times in Einstein's General Relativity,''
Cambridge University Press, 2009,
ISBN 978-1-139-48116-8
%doi:10.1017/CBO9780511635397
%256 citations counted in INSPIRE as of 06 Jun 2026

%\cite{Astorino:2026okd}
\bibitem{Astorino:2026okd}
M.~Astorino,
``Static hairy black hole in 4D general relativity,''
Phys. Rev. D \textbf{113}, no.2, 024047 (2026)
%doi:10.1103/yz86-wc3g
[arXiv:2601.16254 [gr-qc]].
%7 citations counted in INSPIRE as of 06 Jun 2026

%\cite{Ernst:1976mzr}
\bibitem{Ernst:1976mzr}
F.~J.~Ernst,
``Black holes in a magnetic universe,''
J. Math. Phys. \textbf{17}, no.1, 54-56 (1976)
%doi:10.1063/1.522781
%309 citations counted in INSPIRE as of 06 Jun 2026

%\cite{Astorino:2025lih}
\bibitem{Astorino:2025lih}
M.~Astorino,
``Black holes in the external Bertotti-Robinson-Bonnor-Melvin electromagnetic field,''
Phys. Rev. D \textbf{112}, no.10, 104077 (2025)
%doi:10.1103/c5lw-53yd
[arXiv:2508.12908 [gr-qc]].
%15 citations counted in INSPIRE as of 06 Jun 2026

%\cite{Ma:2026otg}
\bibitem{Ma:2026otg}
L.~Ma and H.~Lu,
``Demagnetizing KBR and New Ricci-flat Rotating Metric,''
[arXiv:2605.13954 [gr-qc]].
%1 citations counted in INSPIRE as of 06 Jun 2026

%\cite{Gibbons:2013yq}
\bibitem{Gibbons:2013yq}
G.~W.~Gibbons, A.~H.~Mujtaba and C.~N.~Pope,
``Ergoregions in Magnetised Black Hole Spacetimes,''
Class. Quant. Grav. \textbf{30}, no.12, 125008 (2013)
%doi:10.1088/0264-9381/30/12/125008
[arXiv:1301.3927 [gr-qc]].
%120 citations counted in INSPIRE as of 06 Jun 2026

%\cite{Podolsky:2025tle}
\bibitem{Podolsky:2025tle}
J.~Podolsky and H.~Ovcharenko,
``Kerr Black Hole in a Uniform Bertotti-Robinson Magnetic Field: An Exact Solution,''
Phys. Rev. Lett. \textbf{135}, no.18, 181401 (2025)
%doi:10.1103/rfgv-ybz5
[arXiv:2507.05199 [gr-qc]].
%50 citations counted in INSPIRE as of 06 Jun 2026

%\cite{Ovcharenko:2025cpm}
\bibitem{Ovcharenko:2025cpm}
H.~Ovcharenko and J.~Podolsk{\'y},
``New class of rotating charged black holes with nonaligned electromagnetic field,''
Phys. Rev. D \textbf{112}, no.6, 064076 (2025)
%doi:10.1103/8wkz-th6v
[arXiv:2508.04850 [gr-qc]].
%24 citations counted in INSPIRE as of 07 Jun 2026

%\cite{Herdeiro:2026jem}
\bibitem{Herdeiro:2026jem}
C.~A.~R.~Herdeiro and J.~P.~A.~Novo,
``Vacuum, ma non troppo: hidden matter distribution in symmetry-transformed electrovacuum spacetimes,''
[arXiv:2605.18967 [gr-qc]].
%0 citations counted in INSPIRE as of 11 Jun 2026

\bibitem{planetdata}
Data available at https://ssd.jpl.nasa.gov/planets/approx\_pos.html.

%\cite{Pitjeva:2013xxa}
\bibitem{Pitjeva:2013xxa}
E.~V.~Pitjeva and N.~P.~Pitjev,
``Relativistic effects and dark matter in the Solar system from observations of planets and spacecraft,''
Mon. Not. Roy. Astron. Soc. \textbf{432}, 3431 (2013)
%doi:10.1093/mnras/stt695
[arXiv:1306.3043 [astro-ph.EP]].
%147 citations counted in INSPIRE as of 09 Jun 2026

%\cite{Shapiro:1964uw}
\bibitem{Shapiro:1964uw}
I.~I.~Shapiro,
``Fourth Test of General Relativity,''
Phys. Rev. Lett. \textbf{13}, 789-791 (1964)
%doi:10.1103/PhysRevLett.13.789
%844 citations counted in INSPIRE as of 10 Jun 2026

%\cite{Bertotti:2003rm}
\bibitem{Bertotti:2003rm}
B.~Bertotti, L.~Iess and P.~Tortora,
``A test of general relativity using radio links with the Cassini spacecraft,''
Nature \textbf{425}, 374-376 (2003)
%doi:10.1038/nature01997
%1739 citations counted in INSPIRE as of 10 Jun 2026

%\cite{Reasenberg:1979ey}
\bibitem{Reasenberg:1979ey}
R.~D.~Reasenberg, I.~I.~Shapiro, P.~E.~MacNeil, R.~B.~Goldstein, J.~C.~Breidenthal, J.~P.~Brenkle, D.~L.~Cain, T.~M.~Kaufman, T.~A.~Komarek and A.~I.~Zygielbaum,
``Viking relativity experiment: Verification of signal retardation by solar gravity,''
Astrophys. J. Lett. \textbf{234}, L219-L221 (1979)
%doi:10.1086/183144
%370 citations counted in INSPIRE as of 10 Jun 2026

\bibitem{Lyu2026}
H.-D.~Lyu,  M.~Wang and S.~Li,
``Geodesics and shadows in the spindle-deformed Kerr spacetime,'' in preparation.



%\cite{Li:2020wse}
%\bibitem{Li:2020wse}
%S.~L.~Li, W.~D.~Tan, P.~Wu and H.~Yu,
%``Estimating the final spin of binary black holes merger in STU supergravity,''
%Nucl. Phys. B \textbf{975}, 115665 (2022)
%doi:10.1016/j.nuclphysb.2022.115665
%[arXiv:2003.01957 [gr-qc]].
%2 citations counted in INSPIRE as of 20 Nov 2025




%\cite{Will:2014kxa}
%\bibitem{Will:2014kxa}
%C.~M.~Will,
%``The Confrontation between General Relativity and Experiment,''
%Living Rev. Rel. \textbf{17}, 4 (2014)
%doi:10.12942/lrr-2014-4
%[arXiv:1403.7377 [gr-qc]].
%2771 citations counted in INSPIRE as of 26 Sep 2025

%\cite{Berti:2015itd}
%\bibitem{Berti:2015itd}
%E.~Berti, E.~Barausse, V.~Cardoso, L.~Gualtieri, P.~Pani, U.~Sperhake, L.~C.~Stein, N.~Wex, K.~Yagi and T.~Baker, \textit{et al.}
%``Testing General Relativity with Present and Future Astrophysical Observations,''
%Class. Quant. Grav. \textbf{32}, 243001 (2015)
%doi:10.1088/0264-9381/32/24/243001
%[arXiv:1501.07274 [gr-qc]].
%1495 citations counted in INSPIRE as of 26 Sep 2025







\end{thebibliography}

\end{document}